\documentstyle[12pt]{article}

\begin{document} 

\title{
{\Large 
{\bf Investigations of Azimuthal Asymmetry in 
Semi-Inclusive Leptoproduction }}}

\author{\bf K.A. Oganessyan\thanks{On leave of absence from Yerevan Physics 
Institute, Alikhanian Br.2, 375036 Yerevan, Armenia} \thanks{e-mail: 
kogan@lnf.infn.it} , 
H.R. Avakian\thanks{On leave of absence from Yerevan Physics 
Institute, Alikhanian Br.2, 375036 Yerevan, Armenia} \thanks{e-mail: 
avakian@lnf.infn.it} , 
N. Bianchi\thanks{e-mail: bianchi@lnf.infn.it} , 
P. Di Nezza\thanks{e-mail: dinezza@lnf.infn.it} \\
{\normalsize LNF-INFN, C.P. 13, Enrico Fermi 40, Frascati, Italy}
}

\maketitle

\begin{abstract}
We consider the azimuthal asymmetries in semi-inclusive deep 
inelastic leptoproduction arising due to both perturbative and nonperturbative 
effects at HERMES energies and show that the $k_T^2/Q^2$ order corrections 
to $\langle \cos\phi \rangle$ and $\langle \cos2\phi \rangle$ are 
significant. We also reconsider the results of perturbative effects for 
$\langle \cos\phi \rangle$ at large momentum transfers \cite{A1} using 
the more recent sets of scale-dependent distribution and fragmentation 
functions, which bring up to $18{\%}$ difference in $\langle \cos\phi 
\rangle$. In the same approach we calculate the $\langle \cos2\phi \rangle$ 
as well.       
\end{abstract}

\newpage

The semi-inclusive deep inelastic process $l(k_1) + p(P_1) \to l^{'}(k_2) + 
h (P_2) + X$, where $l$ and $l^{'}$ are charged leptons and $h$ is a observed 
hadron, has been recognized \cite{A2} as an important testing ground 
for QCD. In particular, measurement of the azimuthal angle $\phi$ of the 
detected hadron around the virtual photon direction (Fig.1) provides 
information on the production mechanism. Different mechanisms to generate 
azimuthal asymmetries - $\langle \cos\phi \rangle$ and $\langle \cos2\phi 
\rangle$ have been discussed in the literature. Georgi and Politzer \cite{A2} 
found a negative contribution to  $\langle \cos\phi \rangle$ in the 
first-order in $\alpha_S$ perturbative theory and proposed the measurement 
of this 
quantity as a clean test of QCD. However, partons have nonzero transverse 
momenta ($k_T$) as a consequence of being confined by the strong interactions 
inside hadrons. As Cahn \cite{A3} showed, there is a contribution to $\langle 
\cos\phi \rangle$ from the lowest-order processes due to this intrinsic 
transverse momentum. Therefore the perturbative QCD alone does not describe 
the observed azimuthal angular dependence. In connection with this Chay, 
Ellis and Stirling \cite{A1} combined these perturbative and nonperturbative 
mechanisms and analyzes the quantity $\langle \cos\phi \rangle$ as a function 
of the detected hadron's transverse momentum cutoff $P_C$. 

In this paper we reconsider the results obtained in Ref. \cite{A1} at 
HERMES energies and show that the $k_T^2/Q^2$ order corrections to 
$\langle \cos\phi \rangle$ and $\langle \cos2\phi \rangle$ are significant, 
whereas at $E665$ energies in Fermilab 
\cite{FL} these contributions are less then $10 {\%}$ \cite{A1}. We also 
recalculate the behavior of $\langle \cos\phi \rangle$ in the kinematic 
regime at HERA where perturbative QCD effects should dominate, by using 
the new sets of scale-dependent distribution and fragmentation functions 
which bring up to ${18\%}$ 
difference to quantity $\langle \cos\phi \rangle$ obtained in Ref. \cite{A1}. 
In the same approach the quantity $\langle \cos2\phi \rangle$ is calculated 
as well.   
      
The quantities $\langle \cos\phi \rangle$ and $\langle \cos2\phi \rangle$ are 
defined as 
\begin{equation}
\label{R1}
\langle \cos\phi \rangle = \frac{\int d\sigma^{(0)}\cos\phi + \int 
d\sigma^{(1)}\cos\phi }{\int d\sigma^{(0)} + \int d\sigma^{(1)}},
\end{equation}
\begin{equation}
\label{RR1}
\langle \cos2\phi \rangle = \frac{\int d\sigma^{(0)}\cos2\phi + \int 
d\sigma^{(1)}\cos2\phi }{\int d\sigma^{(0)} + \int d\sigma^{(1)}}, 
\end{equation}       
where $d\sigma^{(0)}$ ($d\sigma^{(1)}$) is the lowest-order (first-order in 
$\alpha_S$) hadronic scattering cross section expressed as 
$$
d\sigma \sim F_{i}(\xi, Q^2) \otimes d\sigma_{ij} \otimes D_{j}(\xi^{'}, 
Q^2),   
$$ 
where $F_{i}(\xi, Q^2)$ is the probability distribution describing an 
$i$-type parton with a fraction $\xi$ of the target momentum, $p_1^{\mu}=
\xi P_1^{\mu}$, $d\sigma_{ij}$ describes the partonic semi-inclusive process 
(Fig.2) and $D_{j}(\xi^{'}, Q^2)$ is the probability distribution for a 
$j$-type parton to fragment producing a hadron with a fraction $\xi^{'}$ of 
the partons momentum, $P_2^{\mu}=\xi^{'} p_2^{\mu}$. In Eqs.(\ref{R1}, 
\ref{RR1}) the integrations are over $P_{2T}$, $\phi$, $x_H$, $y$ and $z_H$. 
These  usual set of kinematic variables are defined as: 
$$
x_H = {Q^2 \over {2(P_1q)}}, \quad y={(P_1q) \over (P_1k_1)}, \quad 
z_H={(P_1P_2) \over (P_1q)},  
$$   
where $q$-momentum of the virtual photon ($Q^2=-q^2$), and the parton variables
$$
x= {x_H \over \xi} = {Q^2 \over {2(p_1q)}}, \quad z = {z_H \over \xi^{'}} = 
{(p_1p_2) \over (p_1q)}.   
$$
The nonperturbative effects are parameterized by Gaussian distributions for 
the intrinsic transverse momenta of both the target (proton) 
and the observed hadron (pion): 
$$
F_i(\xi, Q^2) \to d^2k_T \tilde{F}_i(\xi, \vec{k}_T, Q^2) = d^2k_T 
F_i(\xi, Q^2) f(\vec{k}_T), 
$$
\begin{equation}
\label{R2}
D_j(\xi^{'}, Q^2) \to d^2\rho' \tilde{D}_j(\xi^{'}, \vec{\rho'}, 
Q^2) = d^2\rho' D_j(\xi^{'}, Q^2) d(\vec{\rho'}), 
\end{equation}
where 
$$
f(\vec{k}_T) = {1 \over a^2\pi} e^{-{k_T^2/a^2}}, \quad 
d(\vec{\rho'})={1 \over b^2\pi} e^{-{{\rho'}^2}/b^2},  
$$
and $\vec{\rho'}$ is defined to be perpendicular to the direction of motion 
of the outgoing parton. Then in the center-of-mass frame of the virtual 
photon and the proton (in the limit $M^2/P_1^2 \ll 1, 
P_1^2= Q^2/4x_H(1-x_H)$) the hadron's transverse momentum, perpendicular to 
$\vec{q}$ is given by (for more details see Ref.\cite{A1}) 
\begin{equation}
\label{R3}
\vec{P}_{2T} = \xi^{'} {\vec{k}}_T + \vec{\rho'}-
{(\vec{P}_1 \vec{\rho'}) \over P_1^2} \vec{P}_1, 
\end{equation}
and it's magnitude as 
\begin{equation}
\label{R4}
P_{2T}^2 = {(\xi^{'} \vec{k}_T + \vec{\rho'})}^2 - {4x_H \over {1-x_H}} 
{{(\vec{k}_T \vec{\rho'})}^2 \over Q^2}. 
\end{equation}

If we allow the initial parton to have intrinsic transverse momentum 
$\vec{p}_1=\xi \vec{P}_1 + \vec{k}_T$, the parton cross section at lowest 
order (Fig.2(a)) is modified \cite{A3} to 
$$
\frac{d\sigma_{ij}}{dxdydzdp^2_{2T}d\phi} = {2\pi\alpha^2 \over yQ^2} Q_i^2 
\delta_{ij}\delta(1-x)\delta(1-z) \delta^2(\vec{p_{2T}}-\vec{k}_T)
$$
\begin{equation}
\label{R5}
\left \{1+(1-y)^2+{4p^2_{2T} \over Q^2} (1-y) - {4p_{2T} \over Q} \cos \phi 
(2-y){(1-y)}^{1/2} + {8p^2_{2T} \over Q^2} (1-y) \cos{2\phi} \right \}, 
\end{equation}  
where $Q_i$ is a charge of $i$-type parton. 

Using this parton cross section along with the distribution and fragmentation 
functions of Eq.(\ref{R2}) and observed hadron's transverse momentum defined 
in Eq.(\ref{R3}), one may obtain an explicit expression for the hadronic 
cross section at lowest-order in $\alpha_S$ 
$$
\int d\sigma^{(0)} \cos{\phi} d\phi = -{8\pi^2 \alpha^2 \over Q^2} 
\int dy dx_H dz_H dP^2_{2T} d^2k_Td^2 \rho'{(2-y) {(1-y)}^{1/2} \over y}     
$$
\begin{equation}
\label{R6}
\sum_{j} Q^2_j F_j(x_H,Q^2) D_j(z_H,Q^2){k_T \over Q} f(k_T) d(\rho^{'}) 
\delta^2(\vec{P}_T - \xi^{'} {\vec{k}}_T - \vec{\rho'}+{(\vec{P}_1
\vec{\rho'}) \over P_1^2} \vec{P}_1), 
\end{equation}
$$
\int d\sigma^{(0)} \cos{2\phi}d\phi = {8\pi^2 \alpha^2 \over Q^2} 
\int dy dx_H dz_H dP^2_{2T} d^2k_T d^2 \rho'{ 
{k^2_T \over Q^2} (1-y)}    
$$
\begin{equation}
\label{RR7}
\sum_{j} Q^2_j F_j(x_H,Q^2) D_j(z_H,Q^2)f(k_T) d(\rho') \delta^2(\vec{P}_T - 
\xi^{'} {\vec{k}}_T - \vec{\rho'}+{(\vec{P}_1\vec{\rho'}) 
\over P_1^2} \vec{P}_1),
\end{equation}
$$
\int d\sigma^{(0)} d\phi = {4\pi^2 \alpha^2 \over Q^2} 
\int dy dx_H dz_H dP^2_{2T} d^2k_T d^2 \rho'{\left\{1 + {(1-y)}^2 + 
{4k^2_T \over Q^2} (1-y)\right\}}    
$$
\begin{equation}
\label{R7}
\sum_{j} Q^2_j F_j(x_H,Q^2) D_j(z_H,Q^2)f(k_T) d(\rho') \delta^2(\vec{P}_T - 
\xi^{'} {\vec{k}}_T - \vec{\rho'}+{(\vec{P}_1\vec{\rho'}) 
\over P_1^2} \vec{P}_1),
\end{equation}
where the lower limit of the integrating over $P_{2T}$ is $P_C$ (observed 
hadron's transverse momentum cutoff). 

At large momentum transfers, the intrinsic transverse momenta of the partons 
of a few hundred $MeV$ cannot produce hadrons with larger transverse momenta 
and the nonperturbative effects from $\sigma^{(0)}$ are suppressed. Therefore, 
$\langle \cos\phi \rangle$ and $\langle \cos2\phi \rangle$ are, to a good 
approximations, 
\begin{equation}
\label{R8}
\langle \cos\phi \rangle \approx \frac{\int d\sigma^{(1)}\cos\phi}{\int 
d\sigma^{(1)}}.
\end{equation}
\begin{equation}
\label{RR8}
\langle \cos2\phi \rangle \approx \frac{\int d\sigma^{(1)}\cos2\phi}{\int 
d\sigma^{(1)}}.
\end{equation}
The numerators and denominator of these equations can be written in following 
form 
$$
\int d\sigma^{(1)} {\cos\phi} d{\phi} = \frac{8\alpha_S\alpha^2}{3Q^2} 
\frac{(2-y){(1-y)}^{1/2}}{y} \int_{x_H}^1 {dx \over x} 
\int_{z_H}^1 {dz \over z} \times 
$$
\begin{equation}
\label{R9}
\sum_j Q^2_j \left \{ F_j(\xi,Q^2)AD_j(\xi^{'},Q^2)
+F_j(\xi,Q^2)BD_G(\xi^{'},Q^2)+
F_G(\xi,Q^2)CD_j(\xi^{'},Q^2)\right \},  
\end{equation}
$$
\int d\sigma^{(1)} \cos2\phi d\phi = \frac{8\alpha_S\alpha^2}{3Q^2} 
{{1-y} \over y} \int_{x_H}^1 {dx \over x} 
\int_{z_H}^1 {dz \over z} \times
$$
\begin{equation}
\label{RR9}
\sum_j Q^2_j \left \{ F_j(\xi,Q^2)A^{'}D_j(\xi^{'},Q^2)
+F_j(\xi,Q^2)B^{'}D_G(\xi^{'},Q^2)+
F_G(\xi,Q^2)C^{'}D_j(\xi^{'},Q^2)\right \},
\end{equation}
$$
\int d\sigma^{(1)} d\phi = \frac{4\alpha_S\alpha^2}{3Q^2} 
{1 \over y} \int_{x_H}^1 {dx \over x} 
\int_{z_H}^1 {dz \over z} \times
$$
\begin{equation}
\label{R10}
\sum_j Q^2_j \left \{ F_j(\xi,Q^2)A^{''}D_j(\xi^{'},Q^2)
+F_j(\xi,Q^2)B^{''}D_G(\xi^{'},Q^2)+
F_G(\xi,Q^2)C^{''}D_j(\xi^{'},Q^2)\right \},
\end{equation}
where 
$$
A = - \left \{{{xz} \over {(1-x)(1-z)}}\right\}^{1/2}[xz+(1-x)(1-z)] 
$$
$$
B = \left \{{{x(1-z)} \over {(1-x)z}}\right\}^{1/2}[x(1-z)+(1-x)z]
$$
$$
C = - {3 \over 8}\left \{{{x(1-x)} \over {z(1-z}}\right\}^{1/2}(1-2x)(1-2z)
$$
$$
A^{'} = xz
$$
$$
B^{'} = x(1-z)
$$
$$
C^{'} = {3 \over 4} x(1-x)
$$
$$
A^{''} = [1+{(1-y)}^2] \frac{x^2+z^2}{(1-x)(1-z)}+2y^2(1+xz)+4(1-y)(1+3xz) 
$$
$$
B^{''} = [1+{(1-y)}^2] \frac{x^2+({1-z})^2}{z(1-x)}+2y^2(1+x-xz)+
4(1-y)(1+3x(1-z)) 
$$
$$
C^{''} = {3 \over 8} {\left\{[1+{(1-y)}^2][x^2+({1-x})^2]\frac{z^2+({1-z})^2}
{z(1-z)}+16x(1-y)(1-x)\right \} }
$$
These expressions are identical with previous perturbative results in 
Ref.\cite{A4} and the quantities $A$, $B$ and $C$ and those with primes and 
two primes arise from diagrams Figs.2(b)-2(d) respectively. 
 
Let us consider how $\langle \cos\phi \rangle$ as defined in Eq.(\ref{R1}) 
with $P_{2T}$ cutoff $P_C$, behaves numerically with including both 
leading-order QCD (Eqs.(\ref{R9},\ref{R10})) and 
intrinsic transverse momentum (Eqs.(\ref{R6},\ref{R7})). We use the 
Gl$\ddot{u}$ck, et. al. (GRV) parton distribution functions \cite{GRV} for 
$F_j(\xi, Q^2)$ in Eq.(\ref{R2}) and the scale-dependent Binnewies, et.al. 
(BKK) parameterizations \cite{BKK} for the quark and gluon fragmentation 
functions to charged pions. Our numerical results at HERMES energies: 
$E_l=27.5 GeV$, $Q^2 > 1 GeV^2$, $0.1 < y < 0.85$, $0.02 < x_H< 0.4$ 
and $0.2 < z_H < 1$, presented in Fig. 3 (at these ranges the 
difference in $\langle \cos\phi 
\rangle$ with distribution functions of Ref. \cite{HMS} (HMSR) is less then 
a few percents). In order to make an average over the range of $Q^2$, we 
also (as in Ref. \cite{A1}) use the relation $Q^2=2ME_lx_Hy$, where $M$ is 
the proton mass. The curves correspond to integrating over the same ranges 
with keeping the $k^2_T/Q^2$ term in Eq.(\ref{R4}) (Fig.3 (a)) and with 
neglecting the term of the order $k^2_T/Q^2$ so that $\vec{P}_{2T} = 
\xi^{'}\vec{k}_T+\vec{\rho'}$ approximation (Fig.3 (b)). In both cases we 
take $a=b=0.3 GeV$, which corresponds an average intrinsic transverse momenta 
of $\langle k_T \rangle = \langle p_{T} \rangle = 0.27 GeV$. Note, that  
this choice is arbitrary. The numerical magnitude of $\langle k_T\rangle $ is 
at present rather uncertain and there are not measurements of the unpolarized 
azimuthal asymmetries at HERMES yet. In this respect our aim is  only to show 
the role of the $k^2_T/Q^2$ corrections and their quantitative contributions 
in azimuthal asymmetries at HERMES kinematics (small $Q^2$ and relatively 
large $x$) for some reasonable numerical magnitude of $\langle k_T \rangle $. 
The reason of choice of a small value of mean $k_T$ comes from the fact that 
in the covariant parton model the $k_T$ depends on kinematical variables: the 
small and moderate $Q^2$ (and relatively large $x$) requiring a small mean 
$k_T$. In the same approach we calculate also the angular moment $\langle 
\cos2\phi \rangle$ as defined in Eq.(\ref{RR1}) using the Eqs.(\ref{RR9}, 
\ref{R10}) and Eqs.(\ref{R6},\ref{R7}). The numerical results are illustrated 
in Fig.4. One can see from Figs.3,4 that the contribution of the term 
$k^2_T/Q^2$ to $\langle \cos\phi \rangle$ and to $\langle \cos2\phi \rangle$ 
is significant. Thus, one can conclude that in kinematic regime of HERMES the 
error of $\vec{P}_{2T} = \xi^{'}\vec{k}_T+\vec{\rho'}$ approximation  
(valid to order $k_T/Q$) is rather big. The reason of this is mainly 
conditioned by small $Q^2$ and relatively large $x_H$. Note, that for 
relatively large values of the $\langle k_T \rangle$, $\langle p_{T} 
\rangle$, the magnitudes of the nonperturbative $|\langle \cos\phi \rangle|$ 
and $\langle \cos2\phi \rangle$ increase and the $k^2_T/Q^2$ order 
corrections become more essential. The contributions of perturbative effects 
in this regime are not exceed a few percents.  

Moreover, it is important to mention here that the complete behavior of 
azimuthal distributions may be predicted only after inclusion of higher-twist 
mechanisms, as suggested by Berger \cite{BE}. He considered the case of 
single pion production taking into account pion bound-state effects, 
which generates azimuthal asymmetries with opposite sign respect to 
perturbative QCD and intrinsic transverse momenta effects. More recently 
Brandenburg, et.al. \cite{BKM} reconsidered Berger's mechanism and discussed 
the way of disentangle the effects from those considered above.      

If we now focus to large $Q^2$ values and larger transverse 
momenta for the observed hadrons, the nonperturbative contributions 
are much less important (the contributions of $\sigma^{(0)}$ are negligible). 
In Ref. \cite{A1} the authors estimated $10 {\%}$ theoretical uncertainty, 
due to the indetermination of the distribution and fragmentation functions. 
We recalculate the quantity $\langle \cos\phi \rangle$ by formulae of 
Eq.(\ref{R8}) for the same ranges as in Ref.\cite{A1} using the 
new sets of scale-dependent parton distribution \cite{GRV} and fragmentation 
functions \cite{BKK}. In Fig.5 we also exhibit the result 
of Ref. \cite{A1}, where parton distribution \cite{HMS} and 
scale-independent Segal's fragmentation functions \cite{SH} have been used. 
From Fig.5 one can conclude that those new distribution and fragmentation 
functions bring up to $18{\%}$ difference to quantity $\langle \cos\phi 
\rangle$. This difference arises most probably due to the discrepancy 
between GRV and HMSR distribution functions at small $x$ ($x < 0.1$). 

Fig.6 displays the result for quantity $\langle \cos2\phi \rangle$ calculated 
by formulae of Eq.(\ref{RR8}) in the same range using the recent sets of $Q^2$ 
depending distribution functions. 

In summary, we have investigated the azimuthal asymmetries in semi-inclusive 
deep inelastic leptoproduction arising due to both perturbative and 
nonperturbative effects at HERMES energies. We have showed that due to 
small $Q^2$ and relatively large $x_H$ in that kinematical regime, the 
$k_T^2/Q^2$ order corrections to $\langle \cos\phi \rangle$ and $\langle 
\cos2\phi \rangle$ are significant. At small $Q^2$ (at moderate $Q^2$ as 
well) these quantities are somewhat sensitive to the intrinsic transverse 
momentum, and consequently, the measurement of the azimuthal asymmetries may 
provide a good way to obtain $\langle k_T \rangle$.   

Moreover, we have reconsidered the results of 
perturbative effects for $\langle \cos\phi \rangle$ \cite{A1} in the 
kinematic regime at HERA using the more recent $Q^2$ depending parton 
distribution and fragmentation functions, which bring up to $18{\%}$ 
difference in $\langle \cos\phi \rangle$. In the same approach we have 
calculated the $\langle \cos2\phi \rangle$ as well. 
     
The authors are grateful to A. Kotzinian for useful discussions. 
In addition (K.O) would like to acknowledge D. M$\ddot{u}$ller for 
helpful comments and discussions. 
       
\newpage

\begin{center}
Figure Captions
\end{center}

Fig.1 The definition of the azimuthal angle $\phi$. 

Fig.2 Diagrams contributed to quantity $\langle \cos\phi \rangle$ and 
$\langle \cos2\phi \rangle$ in the zeroth-order in $\alpha_S$ (a), and 
the first-order in $\alpha_S$: (b - d). The dashed line is a gluon.  

Fig.3 $\langle \cos\phi \rangle$ at HERMES energies in the (a) -   
$k^2_T/Q^2$ order, (b) - $k_T/Q$ order. 

Fig.4 $\langle \cos2\phi \rangle$ at HERMES energies in the (a) -   
$k^2_T/Q^2$ order, (b) -  $k_T/Q$ order. 

Fig.5 $\langle \cos\phi \rangle$ at large transfer momentum with using the 
(a) -  parton distribution functions of Ref. \cite{HMS} and fragmentation 
functions of Ref. \cite{SH}, (b) - recent scale-dependent parton distribution 
\cite{GRV} and fragmentation \cite{BKK} functions. The kinematical cuts are 
$Q^2 = 100 GeV^2$, $0.2 < y < 0.8$, $0.05 <x_H< 0.15$ and $0.3 < z_H < 1$.   

Fig.6 $\langle \cos2\phi \rangle$ at large transfer momentum with using the 
recent scale-dependent parton distribution \cite{GRV} and fragmentation 
\cite{BKK} functions. The kinematical cuts are the same as in Fig.5.

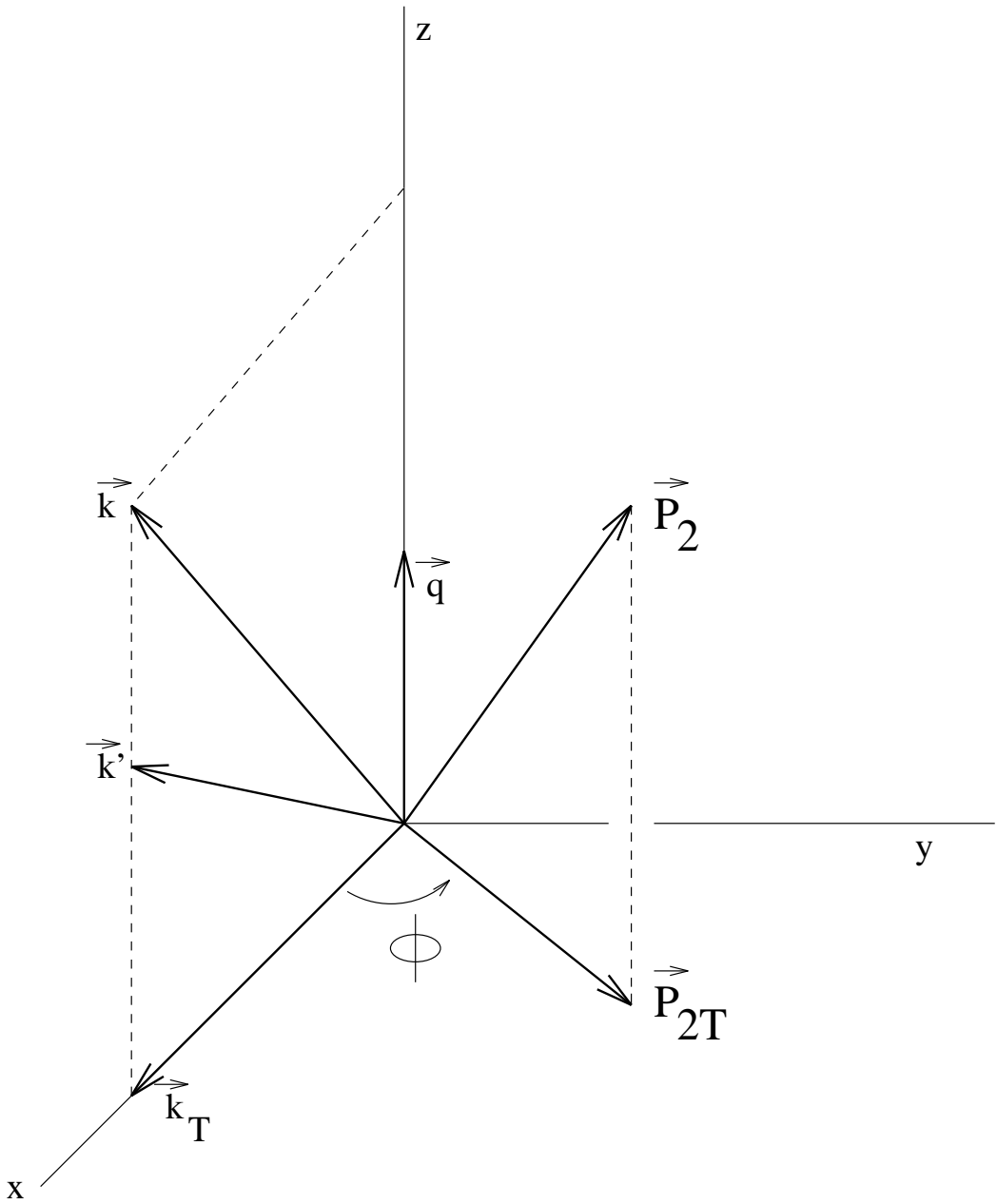
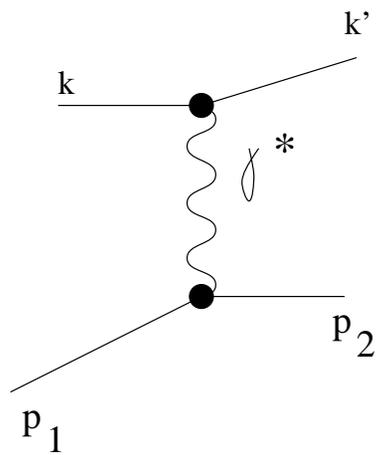
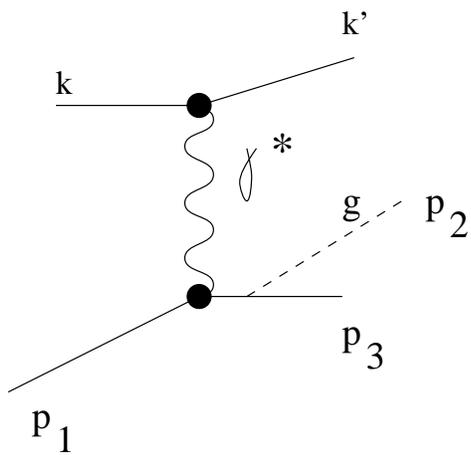
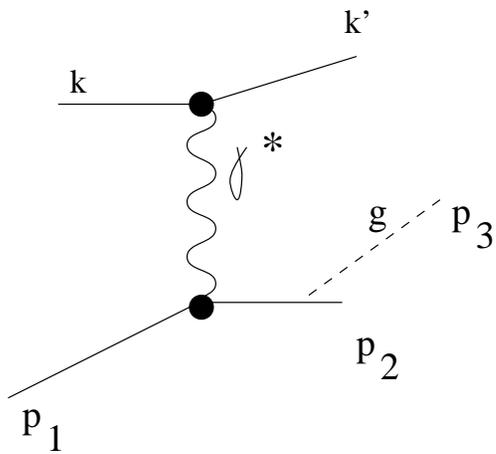
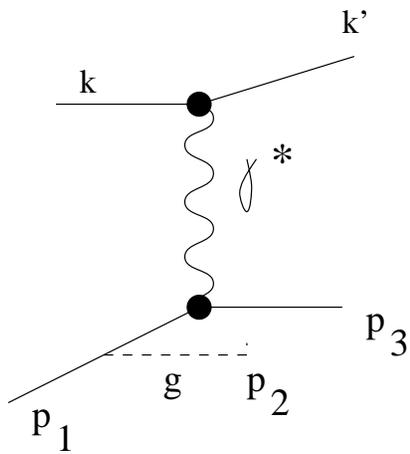
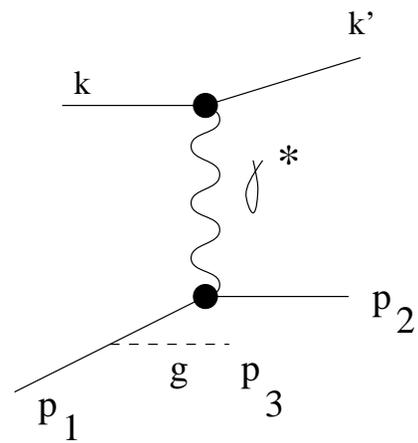
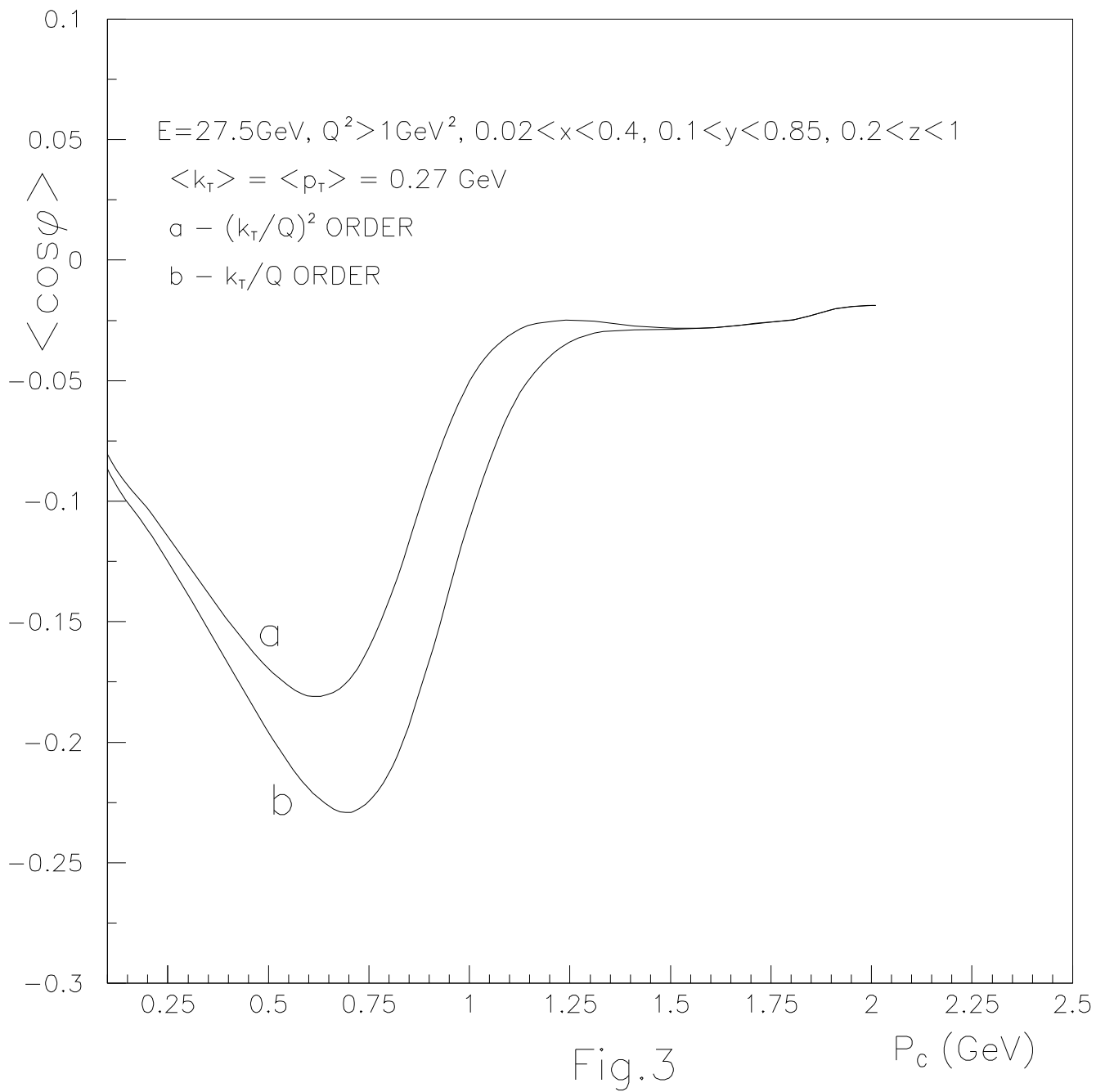
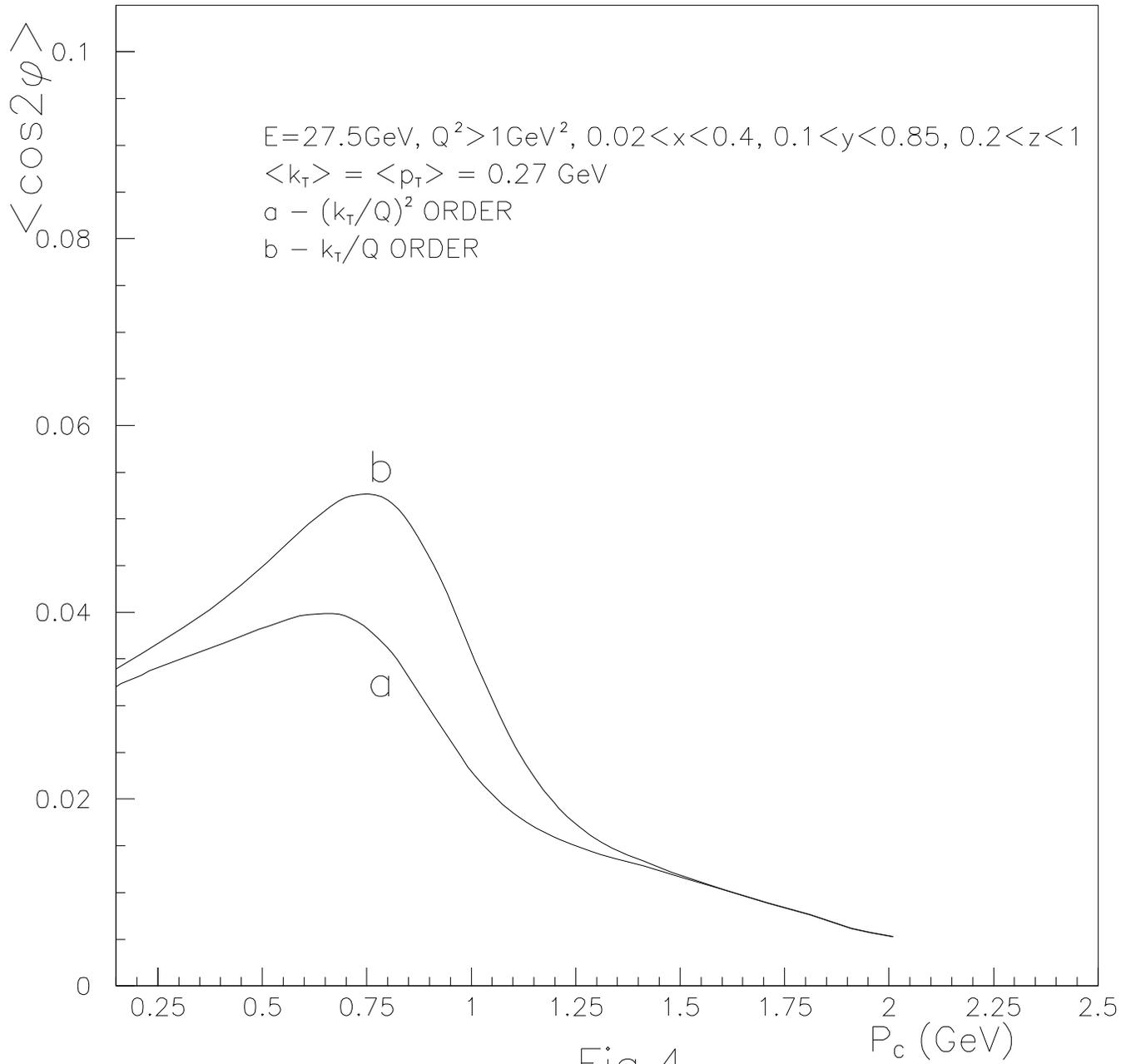
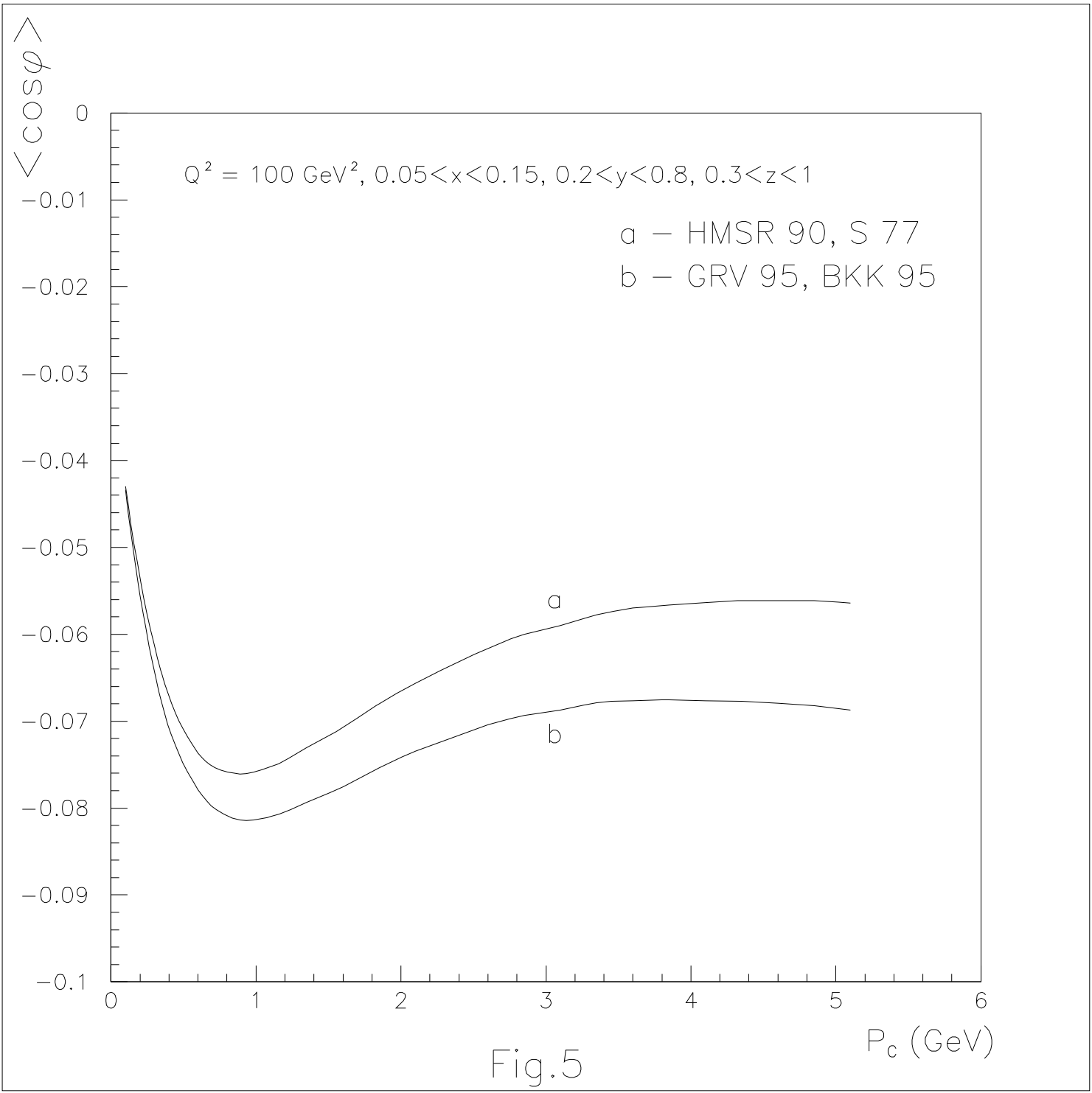
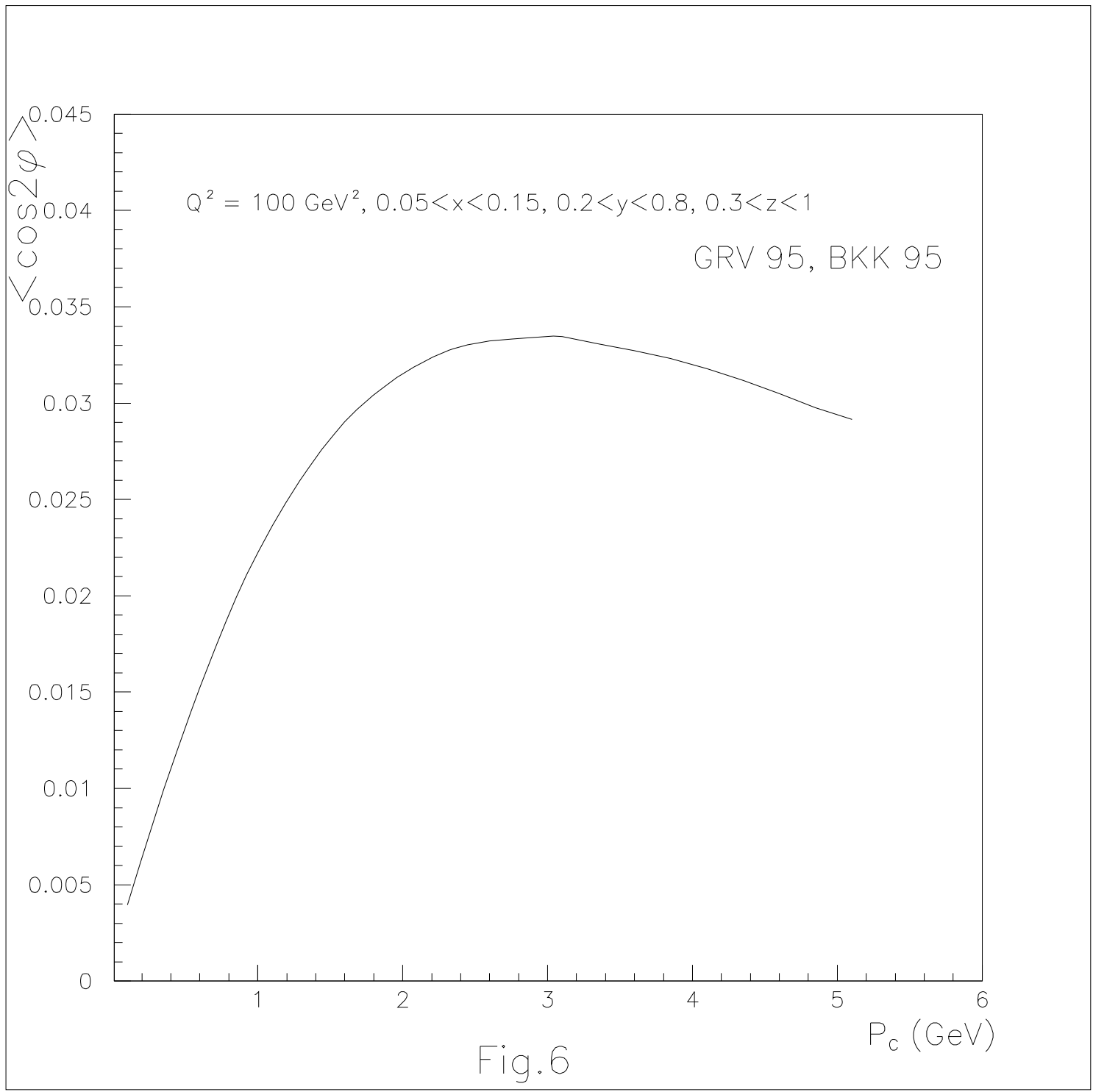
\end{document}